\documentclass[pre,final,showpacs,preprintnumbers,letterpaper]{revtex4}

\usepackage{helvet} 
\usepackage[T1]{fontenc}
\usepackage{bm}
\usepackage{bbm}
\usepackage{amsmath}
\usepackage{amssymb}
\usepackage{graphicx}
\usepackage{units}
\usepackage{geometry}
\newcommand{\tensorhelv}[1]{\fontfamily{phv}\selectfont\textbf{#1}}

\begin{document}



\title{Dynamic ultrasound radiation force in fluids}

\author{Glauber T. Silva}
\email{glauber@tci.ufal.br}
\affiliation{
Departamento de Tecnologia da Informa\c{c}\~ao,
Universidade Federal de Alagoas,
BR 104, km 14, Tabuleiro dos Martins,
Macei\'o, AL, Brazil, 57072-970.}

\author{Shigao Chen, James F. Greenleaf, and Mostafa Fatemi}%
 
\affiliation{
Department of Physiology and Biomedical Engineering,
Mayo Clinic College of Medicine,
200 1st St. SW, Rochester, MN, USA, 55905.}

\date{\today}



\begin{abstract}
The subject of this paper is to present a theory for the dynamic radiation 
force produced by dual-frequency ultrasound beams in lossless and non-dispersive fluids.
A formula for the dynamic radiation force exerted on a three-dimensional object by a 
dual-frequency beam is obtained stemming from the fluid dynamics equations. 
Dependence of the dynamic radiation force to nonlinear effects of the medium is
analyzed.
The developed theory is applied to calculate the dynamic radiation force exerted on
solid elastic (brass and stainless steel) spheres by a low-amplitude dual-frequency 
plane wave.  
The dynamic radiation force is calculated by solving the dual-frequency
plane wave scattering problem for solid elastic spheres.
Results show that the dynamic radiation force on the analized spheres presents 
fluctuations similar to those present in the static radiation force.
Furthermore, analysis of the static and the dynamic radiation forces on the stainless steel 
sphere shows that they have approximately the same magnitude.
\end{abstract}

\maketitle


\section{Introduction}
It is known that acoustic waves carry momentum. 
When an acoustic wave strikes an object, part of its momentum is transferred to the 
object giving rise to the acoustic radiation force phenomenon~\cite{borgnis53b,beyer78a,torr84a,lee93a}.
Acoustic radiation force has found practical importance in many applications. 
For example, measure the power output of transducers in medical ultrasound 
machines~\cite{carson78a}, ultrasound radiometer~\cite{chivers82a}, liquid drops 
levitation~\cite{lee91a}, and motion of gas bubbles in liquids~\cite{crum83a}.

In these applications, the radiation force is static, being generated by a continuous-wave 
(CW) ultrasound beam. 
Time-dependent (dynamic) ultrasound radiation force can also be produced by
amplitude-modulated (AM) or pulsed ultrasound beams.
In general, an AM beam produces a harmonic radiation force at the modulation frequency,
while a pulsed beam generates a pulsed radiation force.
Dynamic radiation force has been applied for measuring the ultrasound power 
of transducers using a disk target~\cite{sivian28a} or a shaped-wedge
vane~\cite{greenspan78a}, and determining  ultrasound absorption coefficient in 
liquids~\cite{mcnamara52a}. 

In recent years, dynamic ultrasound radiation force has become of practical importance 
in elastography; specifically in the following imaging techniques: 
\begin{itemize}
\item 
Acoustic radiation force impulse imaging (ARFI) uses pulsed ultrasound 
radiation force to produce displacement in tissue which is detected to form an image of the
tissue~\cite{nightingale02a}.
\item 
Shear wave elasticity imaging (SWEI) images tissue properties by detecting shear acoustic waves induced by the harmonic ultrasound radiation force produced by an AM ultrasound 
beam~\cite{sarvazyan98a}. 
\item 
Vibro-acoustography maps the mechanical response of an object to a harmonic 
ultrasound radiation force produced by two overlapping CW ultrasound beams 
whose frequencies are slightly different~\cite{fatemi98a,fatemi99a}. 
In this context, it has been demonstrated that viscoelastic properties of gel 
phantoms can be accurately determined by measuring the amplitude of vibration induced by
the dynamic ultrasound radiation force on a small sphere embedded in the 
medium~\cite{chen02a}.
\end{itemize}

\begin{figure}[t]
  \centering
  \includegraphics[width=5.5in]{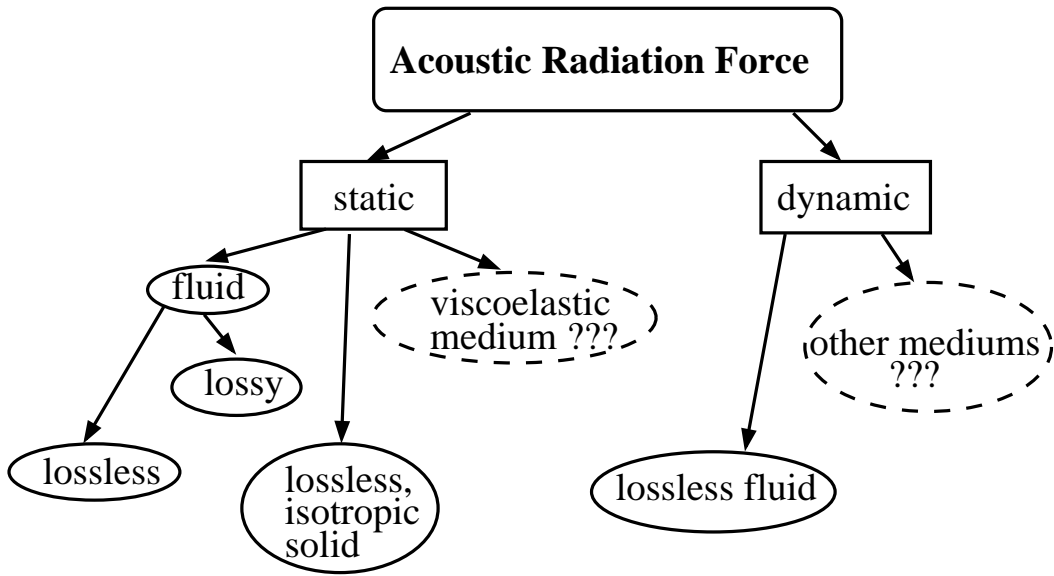}
  \caption
  {Diagram showing the theory of acoustic radiation force in different propagating mediums.
  Dashed ellipses show the lack of theory for acoustic radiation force.}
  \label{fig:diagram}
\end{figure}

Lord Rayleigh was the first to propose a theory for acoustic radiation force in lossless 
fluids due to compressional waves~\cite{rayleigh02a}. 
Static radiation force in lossy fluids has been studied by Jiang 
\textit{et al.}~\cite{jiang96a}, Doinikov~\cite{doinikov96a}, and Danilov 
\textit{et al.}~\cite{danilov00a} 
A study of the static radiation force in lossless isotropic elastic solid can be seen in Ref~\cite{cantrell84a}.
Usually the radiation force exerted on an object target by a CW ultrasound can be obtained 
by solving the vector surface integral of the radiation-stress tensor (defined as
the time-average of the wave momentum flux) over a surface enclosing the object. 
The radiation-stress is obtained in terms of the incident beam and the scattered field 
by the object.
Several authors derived the static radiation force, by solving the scattering 
problem of CW plane waves by spherical or cylindrical 
objects~\cite{king34a,yosioka55a,westervelt57a,maidanik57a,gorkov62a,hasegawa69a,hasegawa88a}. 
Most studies realized for the dynamic radiation force have been focused
on finding applications to it.
No theoretical efforts to tackle the problem of dynamic ultrasound radiation 
force exerted on an embedded object in a medium have been made whatsoever.
Figure~\ref{fig:diagram} depicts the theoretical realm of acoustic radiation force. 
This figure includes the contribution of this paper: dynamic radiation force in lossless fluids.
Dashed ellipses show the lack of theoretical models for acoustic radiation force.

The increasing applications of dynamic radiation force of ultrasound in elastography techniques prompted us to develop a method to calculate this force.
Here, we present a theory of dynamic ultrasound radiation force exerted on arbitrary shaped 
three-dimensional objects. 
The theory is restricted to the radiation force produce by dual-frequency CW ultrasound
waves in lossless fluids.
In what follows, we briefly discuss the dynamic equations of lossless fluids in 
Sec.~\ref{sec:ideal_fluids}. 
In Sec.~\ref{sec:radiation_force}, we present a theory of ultrasound radiation
force. 
We obtain a formula for the dynamic ultrasound radiation force exerted
on an object by a dual-frequency CW ultrasound wave.
The formula is given in terms of a vector surface integral of the wave velocity potential over the object's surface.
Explicit dependence of the dynamic radiation force with the medium nonlinearity is
pointed out.
In Sec.~\ref{sec:solid_sphere}, we apply the theory to calculate the dynamic ultrasound 
radiation force exerted on solid elastic (brass and stainless steel) spheres. 
Finally, in Sec.~\ref{sec:discussion} we summarize the main results of this paper.

\section{Theory}
\subsection{Dynamics of lossless fluids}
\label{sec:ideal_fluids}
Consider a homogeneous isotropic fluid in which thermal conductivity and viscosity are 
neglected. 
This corresponds to the so-called ideal fluid.
The medium is characterized by the following acoustic fields: pressure $p$, density $\rho$,
 and particle velocity $\mathbf{v}$. 
Here all acoustic fields are functions of the position vector
$\mathbf{r} \in \mathbb{R}^3$ and time $t \in \mathbb{R}$.
In an initial state without sound perturbation these quantities assume
constant values given by $p=p_{0}$, $\rho = \rho_{0}$, and $\mathbf{v}=0$. 
The quantity $p-p_{0}$ is the excess of pressure in the medium. 
The equations that describe the dynamic of ideal fluids can be derived from conservation 
principles for mass, momentum, and thermodynamic equilibrium. 
These equations, neglecting effects of gravity, are presented here as 
follows~\cite{landau87a}:
\begin{align}
\label{mass_conservation}
\frac{\partial \rho}{\partial t}+\nabla \cdot (\rho \mathbf{v}) &= 0,\\
\label{euler}
\rho \frac{d \mathbf{v}}{d t} &= - \nabla p,\\
\label{state}
p &= p_0 \left( \frac{\rho}{\rho_0} \right)^{(1+B/A)},
\end{align}

where $\nabla$ is the gradient operator, the symbol $\cdot$ stands for the scalar product,
and $B/A$ is the Fox-Wallace parameter~\cite{fox54a}
which characterizes the nonlinearity of the fluid.
Equations~(\ref{mass_conservation})-(\ref{state}) form a system of nonlinear 
partial differential equations that gives a full description of the wave 
propagation in the fluid. 
The conservation of fluid mass is represented by Eq.~(\ref{mass_conservation}).
Equation~(\ref{euler}) is a version of the Newton's second-law in fluid dynamics.
Equation~(\ref{state}) is an adiabatic equation of state known as Tait's equation. 

A lossless fluid is irrotational. 
Thus, according to the Helmholtz vector theorem, the particle velocity can be expressed in 
terms of the velocity potential function $\phi$ as 
\begin{equation}
\label{velocity_phi}
\mathbf{v}= -\nabla \phi.
\end{equation} 
The velocity potential can be expanded as a sum of a linear term
and higher-order contributions as follows
\begin{equation}
\label{phi_expansion}
\phi = \phi^{(1)} + \phi^{(2)} + ...
\end{equation} 
where $\phi^{(1)}$ and $\phi^{(2)}$ are the linear and the second-order
velocity potentials, respectively. 
In terms of the linear velocity potential, the linear pressure and velocity 
fields are given by
\begin{align}
\label{p1}
p^{(1)} &= \rho_0 \frac{\partial \phi^{(1)}}{\partial t}, \\
\label{v1}
\mathbf{v}^{(1)} &= - \nabla {\phi^{(1)}}.
\end{align}

\subsection{Instantaneous net force}
\label{sec:radiation_force}

A volume element in a fluid is subject to a stress caused by the
sound wave propagation throughout it.
Stresses caused by sound perturbation in the fluid should be described by Eq.~(\ref{euler}). 

\begin{figure}[t]
  \centering
  \includegraphics[width=5.5in]{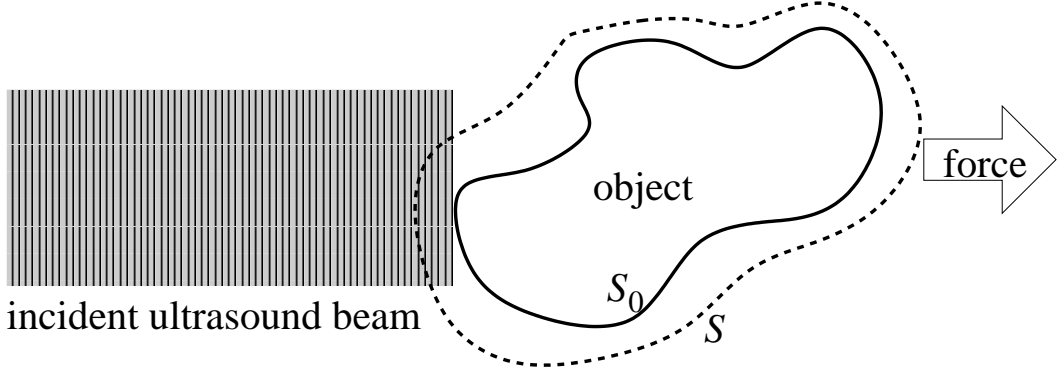}
  \caption
  {Net force exerted on an object by an ultrasound beam. 
   The dotted contour depicts changes on the object's surface.}
  \label{fig:surfaces}
\end{figure}

Consider an ultrasound beam striking a homogeneous object of finite extension and
surface $S_0$ at rest. 
As the ultrasound field hits the object, its surface may be deformed and dislocated. 
We denote the object's surface at the time $t$ by $S$.
Fig.~\ref{fig:surfaces} depicts the interaction between the incident ultrasound wave 
and the object target. 

The instantaneous net force $\mathbf{f}$ acting on the object is obtained by integrating 
Eq.~(\ref{euler}) on the object's volume.
Since the ambient pressure $p_0$ does not contribute to the net force on the object,
we can replace $p$ by $p-p_0$ in Eq.~(\ref{euler}).
Hence, integrating Eq.~(\ref{euler}) on the object's volume and using the Gauss' integral 
theorem, we obtain
\begin{equation}
\label{f_instantaneous}
\mathbf{f} = - \int_S (p-p_0) \mathbf{n}dS,
\end{equation} 
where $\mathbf{n}$ is the outward normal unit-vector of the integration surface.

Radiation force is a phenomenon that depends on the interaction of second-order
acoustic fields with the object target.
To avoid the integration of $p-p_0$ over the time-dependent object's 
surface $S$, we need to find $p-p_0$ up to second-order.
The idea is to replace $S$ by $S_0$ for second-order integrands in 
Eq.~(\ref{f_instantaneous}). 
For first-order integrands, we should keep $S$ and analyze the contribution
of the integral to the radiation force. 
From Eqs.~(\ref{mass_conservation}) - (\ref{state}), one can show that 
the second-order excess of pressure is given by~\cite{olsen58a}
\begin{equation}
\label{excess_2nd}
p - p_0 =  \rho_0 \left( \frac{\partial \phi^{(1)}}{\partial t} 
+ \frac{\partial \phi^{(2)}}{\partial t} \right )
+ \frac{p^{(1)2}}{2\rho _{0}c^{2}_{0}}
- \frac{\rho_0 \mathbf{v}^{(1)}\cdot \mathbf{v}^{(1)}}{2},
\end{equation}
where $c_0$ is the small-amplitude speed of sound.
Substituting Eq.~(\ref{excess_2nd}) into Eq.~(\ref{f_instantaneous}) and
holding terms up to second-order, we find
\begin{equation}
\label{f_inst2}
\mathbf{f} = -\rho_0 \int_S \frac{\partial \phi^{(1)}}{\partial t} \mathbf{n}dS 
- \rho_0 \int_{S_0} \frac{\partial \phi^{(2)}}{\partial t} \mathbf{n} dS
- \frac{1}{2\rho _{0}c^{2}_{0}} \int_{S_0} p^{(1)2} \mathbf{n} dS
+ \frac{\rho_0}{2} \int_{S_0} (\mathbf{v}^{(1)}\cdot \mathbf{v}^{(1)}) \mathbf{n} dS.
\end{equation}
By using the relation~\cite{yosioka55a}
\begin{equation*}
\int_S \frac{\partial \phi^{(1)}}{\partial t} \mathbf{n}dS 
= \frac{d}{dt} \int_S \phi^{(1)} \mathbf{n} dS 
+ \int_S (\mathbf{n} \cdot \mathbf{v}^{(1)}) \mathbf{v}^{(1)} dS
\end{equation*}

in Eq.~(\ref{f_inst2}) and again keeping up to the second-order terms, we get
\begin{equation}
\label{f_inst3}
\mathbf{f} = -\rho_0 \frac{d}{dt} \left( \int_S \phi^{(1)} \mathbf{n} dS  
+ \int_{S_0} \phi^{(2)} \mathbf{n} dS \right)
- \int_{S_0} \mathbf{n} \cdot \tensorhelv{T} dS.
\end{equation}
We call the quantity $\tensorhelv{T}$ as the radiation-stress tensor. 
It is given by

\begin{equation}
\label{rad_stress}
\tensorhelv{T} \equiv \left [ \frac{p^{(1)2}}{2\rho _{0}c^{2}_{0}}-
\frac{\rho_0(\mathbf{v}^{(1)}\cdot \mathbf{v}^{(1)})}{2} \right ] \tensorhelv{I}
+ \rho_0 \mathbf{v}^{(1)}\mathbf{v}^{(1)},
\end{equation}
where $\tensorhelv{I}$ is the $3\times3$-unit matrix and $\rho_0\mathbf{v}^{(1)}\mathbf{v}^{(1)}$
is a dyad~\cite{mase70a} known as the Reynolds' stress. 
We have written down Eq.~(\ref{rad_stress}) using the identities 
$\mathbf{n} \cdot \mathbf{v}^{(1)}\mathbf{v}^{(1)} 
= (\mathbf{n} \cdot \mathbf{v}^{(1)}) \mathbf{v}^{(1)}$
and
$\mathbf{n} \cdot \tensorhelv{I} = \mathbf{n}.$

Consider $g$ as a function of time. 
The Fourier transform of $g$ is defined as 
\begin{equation*}
\hat{g}(\omega) = \mathcal F[g] \equiv \int_{-\infty}^{+\infty} g(t) e^{-j\omega t} dt,
\end{equation*}
where $\omega$ (angular frequency) is the reciprocal variable of $t$
and $j$ is the imaginary-unit.
To analyze the frequency components present in the net of force, we take the Fourier 
transform of Eq.~(\ref{f_inst3}) as follows
\begin{equation}
\label{dynamic_force1}
\hat{\mathbf{f}}(\omega) =  
- j \omega \rho_0  \mathcal{F}\left[\int_{S} \phi^{(1)} \mathbf{n} dS \right]  
- j \omega \rho_0 \int_{S_0}  \mathcal{F}\left[\phi^{(2)}\right] \mathbf{n} dS
- \int_{S_0}  \mathbf{n} \cdot \mathcal{F}[\tensorhelv{T} ] dS.
\end{equation}
This equation gives a description of the frequency spectrum of the net of force
acting on the object. 
We shall use this equation to calculate the static and the dynamic radiation forces.

\subsection{Static radiation force}
The static component of the net force $\mathbf{f}$ is commonly known as acoustic 
radiation force.
Here, we call it static radiation force.
Acoustic radiation force has been studied as a time averaged force.
The rule of the time-average is to isolate the static component $(\omega = 0)$ of the net
force acting on the object. 

Consider an incident ultrasound wave, periodic in time, striking the object target.
The static component of the radiation force corresponds to $\omega = 0$ 
in Eq.~(\ref{dynamic_force1}). 
The first two integrals in the right-hand side of this equation become zero.
Therefore, the static radiation force is 
$$
\mathbf{f}_s = - \int_{S_0} \mathbf{n} \cdot \mathcal{F}[\tensorhelv{T} ]_{\omega = 0} dS.
$$
Recognizing that the time-average of $\tensorhelv{T}$ over a long time interval is
$\langle \tensorhelv{T} \rangle = \mathcal{F}[\tensorhelv{T} ]_{\omega = 0}$,
we get
\begin{equation}
\label{static_f_fourier}
\mathbf{f}_s = -\int_{S_0} \mathbf{n} \cdot \langle \tensorhelv{T} \rangle dS.
\end{equation}
Note that $\mathbf{f}_s$ is a real quantity.

The static radiation force can be understood as follows.
The incident ultrasound beam is scattered by the object.
The static radiation force is the time averaged rate of the momentum change due to the scattering by the object. 

The time-average of the radiation-stress is a zero-divergence quantity, i.e.,
$\nabla \cdot \langle \tensorhelv{T} \rangle = 0$~\cite{lee93a}. 
This means that no static radiation force is present in an ideal fluid if there is no target.
Consequently, steady streaming does not happen in lossless fluids~\cite{lighthill78a}.

\subsection{Dynamic ultrasound radiation force}
In this section, we concentrate on the dynamic radiation force produced by an 
amplitude-modulated (AM) ultrasound beam. 
We chose the dynamic ultrasound force produced by AM ultrasound
waves because of its importance in elastography; specifically in SWEI and 
vibro-acoustography. 
An AM ultrasound wave is described by a carrier $f_0$ and modulation $\Delta f$ 
frequencies. 
Usually the carrier frequency is much larger than the modulation frequency. 

Here, the AM ultrasound beam is produced by two overlapping CW ultrasound beams.
We call this beam a ``dual-frequency ultrasound beam''. 
The frequencies of this beam are $f_a = f_0 + \Delta f/2$ and $f_b = f_0 - 
\Delta f/2$.
The corresponding angular frequencies of the beam are 
$\omega_a = 2\pi f_a$ and $\omega_b = 2\pi f_b$.
The frequencies $f_0$ and $\Delta f$ are also called the center and the difference 
frequencies, respectively.

When an object is placed in the wave path, the dual-frequency ultrasound beam is scattered
by the object.
The first-order velocity potential in the medium is given by
\begin{equation}
\label{AM_beam}
\phi^{(1)} = \Re \left\{ \hat{\phi}_a e^{j \omega_a t} + \hat{\phi}_b e^{j \omega_b t}
\right\},
\end{equation}
where $\Re\{\cdot\}$ is the real-part of a complex quantity.
The functions $\hat{\phi}_a$ and $\hat{\phi}_b$ are spatial complex amplitudes of
each frequency component of the wave.
The amplitude functions in Eq.~(\ref{AM_beam}) are given as the sum of the velocity 
potential amplitudes of the incident and scattered waves. 

The radiation force exerted on the object by the dual-frequency beam includes the following components: a static component $(\omega = 0)$, a component at the difference frequency 
$\Delta \omega$, and high-frequency components at $2 \omega_a$, $2 \omega_b$,
and $\omega_a +\omega_b$.
In this paper, the component at $\Delta \omega$ is called the dynamic radiation force.

To obtain the static and the dynamic radiation force produced by the dual-frequency 
ultrasound beam, we calculate the Fourier transform of the radiation stress $\tensorhelv{T}$
up to the difference frequency $\Delta \omega$.
Substituting Eq.~(\ref{AM_beam}) into Eq.~(\ref{rad_stress}), through 
Eqs.~(\ref{p1}) and (\ref{v1}), and taking the Fourier transform of the result, we obtain
the static, $\langle \tensorhelv{T} \rangle$, and the dynamic, $\hat{\tensorhelv{T}}_{\Delta \omega}$, radiation-stresses as follows
\begin{align}
\langle \tensorhelv{T} \rangle &= \langle \tensorhelv{T}_a \rangle + \langle \tensorhelv{T}_b \rangle,\\
\label{T_ab}
\hat{\tensorhelv{T}}_{\Delta \omega} &=  \frac{\rho_0}{2} \left[ \left ( k_a k_b\hat{\phi}_a\hat{\phi}_b^* 
- \nabla\hat{\phi}_a \cdot \nabla \hat{\phi}_b^* \right ) \tensorhelv{I} 
+ \nabla\hat{\phi}_a \nabla\hat{\phi}_b^* 
+ \nabla\hat{\phi}_b^* \nabla\hat{\phi}_a \right]. 
\end{align}
The quantities $\langle \tensorhelv{T}_a \rangle$ and $\langle \tensorhelv{T}_b \rangle$ are 
the averaged radiation-stress of each frequency component of the dual-frequency wave
and $k_a = \omega_a/c_0$ and $k_b = \omega_b / c_0$.
Explicitly, the averaged radiation-stresses $\langle \tensorhelv{T}_a \rangle$ 
and $\langle \tensorhelv{T}_b \rangle$ are given by 
\begin{equation}
\langle \tensorhelv{T}_m \rangle = \frac{\rho_0}{4} 
\left( k_m^2 |\hat{\phi}_m|^2 - |\nabla \hat{\phi}_m|^2 \right) \tensorhelv{I}
+ \frac{\rho_0}{2} (\nabla \hat{\phi}_m \nabla \hat{\phi}_m^* ), 
\quad m = a,b.\\
\end{equation}
Now both the static and dynamic radiation force can be obtained.
According to Eq.~(\ref{static_f_fourier}) the static radiation force is
\begin{equation}
\label{tot_static}
\mathbf{f}_s = -\int_{S_0} \mathbf{n} \cdot (\langle \tensorhelv{T}_a \rangle 
+ \langle \tensorhelv{T}_b \rangle ) dS.
\end{equation}
Based on Eq.~(\ref{dynamic_force1}) the amplitude (in complex notation) of the dynamic 
radiation force at $\Delta \omega$ is
\begin{equation}
\label{final_force}
\hat{\mathbf{f}}_{\Delta \omega} =  
- j \rho_0 \Delta \omega \left \{
\mathcal{F}\left[ \int_S \phi^{(1)} \mathbf{n} dS \right]_{\omega =\Delta\omega}
+ \int_{S_0} \hat{\phi}^{(2)}_{\Delta \omega} \mathbf{n} dS \right \}
-\int_{S_0} \mathbf{n} \cdot \hat{\tensorhelv{T}}_{\Delta \omega} dS. 
\end{equation} 
where $\hat{\phi}^{(2)}_{\Delta \omega} = 
\mathcal{F} [ \phi^{(2)}]_{\omega= \Delta \omega}$.

Let us analyze the contribution of the first integral in the right-hand side of Eq.~(\ref{final_force}). 
By using the Gauss' integral theorem, this integral can be expressed
as
\begin{equation*}
\int_S \phi^{(1)} \mathbf{n} dS =
-\int_{V_0} \mathbf{v}^{(1)} dV - \int_{\delta V} \mathbf{v}^{(1)} dV,
\end{equation*}
where $V_0$ is volume of the object at rest and $\delta V$ is the volume variation 
of the object induced by the excess of pressure at the time $t$ (see Fig.~\ref{fig:surfaces}).
The first integral in this equation does not have any harmonic term at the difference frequency
$\Delta \omega$, thus, by taking the Fourier transform of it and isolating the frequency component $\omega = \Delta \omega$, we obtain
\begin{equation*}
\mathcal{F}\left[ \int_S \phi^{(1)} \mathbf{n} dS \right]_{\omega =\Delta\omega} = 
-\mathcal{F}\left[ \int_{\delta V} \mathbf{v}^{(1)} dV \right]_{\omega =\Delta\omega}.
\end{equation*}
Inasmuch as the particle velocity is a limited and continuous function inside the volume 
$\delta V$, there exists a point $\mathbf{r}_0 \in \delta V$ such that~\cite{elon81a}
$
\int_{\delta V} \mathbf{v}^{(1)} dV = \delta V \mathbf{v}^{(1)}(\mathbf{r}_0,t).
$ 
Thus, the amplitude of the dynamic radiation force of the first term in 
Eq.~(\ref{final_force}) is
\begin{equation}
\label{F_phi_1}
F_0 = -j\rho_0 \Delta \omega \mathcal{F}\left[ \int_S \phi^{(1)} \mathbf{n} dS \right]_{\omega 
=\Delta\omega} =
j \Delta \omega \mathcal{F}\left[\delta M \mathbf{v}_0^{(1)}\right]_{\omega 
=\Delta\omega},
\end{equation}
where $\mathbf{v}_0^{(1)} = \mathbf{v}^{(1)}(\mathbf{r}_0,t)$ and $\delta M = \rho_0 \delta V$ corresponds to the fluid mass variation at the time $t$ caused
by the object vibration.
Therefore, the dynamic radiation force can be written as
\begin{eqnarray}
\nonumber
\hat{\mathbf{f}}_{\Delta \omega}  &=&   
 j \Delta \omega \mathcal{F}\left[\delta M \mathbf{v}_0^{(1)}\right]_{\omega =\Delta\omega}
 -j \Delta \omega\rho_0\int_{S_0} \hat{\phi}^{(2)}_{\Delta \omega} \mathbf{n}\, dS\mbox{}\\
&&\mbox{} -
\frac{\rho_0}{2} \int_{S_0} \left[ \left ( k_ak_b \hat{\phi}_a \hat{\phi}_b^* 
- \nabla \hat{\phi}_a \cdot \nabla \hat{\phi}_b^* \right ) \mathbf{n} 
+  (\mathbf{n} \cdot \nabla\hat{\phi}_a) \nabla\hat{\phi}_b^* 
+ (\mathbf{n} \cdot \nabla\hat{\phi}_b^*) \nabla\hat{\phi}_a  \right] dS. 
\label{f_force2}
\end{eqnarray}

Thus, the dynamic radiation force exerted on the object target comes from three different
interactions between the ultrasound field and the object.
The first term in the right-hand side of Eq.~(\ref{f_force2}) corresponds to momentum
rate exchange due to fluid mass variation caused by the object vibration. 
The next term in this equation accounts for the interaction of
the second-order velocity potential with the object target. 
None of these terms are present on the static radiation force formula~(\ref{tot_static}). 
The last term in Eq.~(\ref{f_force2}) is related to the interaction of the radiation-stress 
and the object.
We shall apply Eq.~(\ref{f_force2}) to calculate the dynamic radiation force on
a solid small sphere. 

In time-domain, the dynamic radiation force is given as the inverse Fourier transform 
of $\hat{\mathbf{f}}_{\Delta \omega}$. 
Therefore, the dynamic radiation force exerted on a object is given by
\begin{equation}
\label{rad_force_time_domain}
\mathbf{f}_{\Delta \omega}(t) = 
\Re \left\{ \hat{\mathbf{f}}_{\Delta \omega}e^{j \Delta \omega t} \right\}.
\end{equation}
Note that if $\Delta \omega = 0$ and $\hat{\phi}_a = \hat{\phi}_b$,
we have $\hat{\mathbf{f}}_{\Delta \omega} = \mathbf{f}_s$. 
Consequently, the magnitude of both dynamic and static radiation forces becomes
equal when  $\Delta \omega=0$.
  
\section{Dynamic radiation force on a solid sphere}
\label{sec:solid_sphere}

\subsection{Linear ultrasound scattering}
\label{subsec:us_scat}
Consider a collimated dual-frequency plane wave described by Eq.~(\ref{AM_beam}) impinging 
on a solid elastic sphere of radius $r_0$ localized at the origin of the coordinate system. 
The beam propagates along the $z$-axis.
The sphere is characterized by three parameters: density $\rho_1$, compressional 
wave speed $c_c$, and shear wave speed $c_s$. 
The amplitude functions of the incident plus the scattered waves are given,
in spherical coordinates, by~\cite{faran51a}
\begin{equation}
\label{tot_phi}
\hat{\phi}_m = A_0 \sum_{n=0}^{\infty}(2n+1)(-j)^n [j_n(k_m r) + S_{m,n} h^{(2)}_n(k_m r)] 
P_n(\cos \theta), 
\quad m=a,b.
\end{equation}
where $A_0$ is the magnitude of the incident wave, $k_m=\omega_m/c_0$,
$j_n$ is the spherical Bessel function, $S_{m,n}$ is the scattering function 
determined by boundary conditions, and $h_n^{(2)}$ is the spherical Hankel 
function of second-kind.
The scattering function is given by
\begin{equation}
\label{S_mn}
S_{m,n} = -\frac{D_{m,n}j_n(x_m) - x_m j'_n(x_m)}
{D_{m,n}h^{(2)}_n(x_m) - x_m {h^{(2)}_n}'(x_m)}, \quad m=a,b,
\end{equation}  
where the prime symbol means derivative with respect to the function's 
argument and $x_m=k_m r_0$. 
The coefficient $D_{m,n}$ is given by
\begin{equation}
\begin{split}
\label{D_mn}
D_{m,n} 
& = \frac{\rho_0 (x_{m,s})^2}{2 \rho_1}
\left[
\frac{n j_n(x_{m,c}) - x_{m,c} j_{n+1}(x_{m,c})}{(n-1)j_n(x_{m,c})
- x_{m,c} j_{n+1}(x_{m,c})} - \frac{2n(n+1)j_n(x_{m,s})}
{(2n^2-x_{m,s}^2 - 2)j_n(x_{m,s}) + 2 x_{m,s} j_{n+1}(x_{m,s})} \right] \\
& \mbox{} \times
\left[
\frac{(x_{m,s}^2/2 - n^2 + n) j_n(x_{m,c}) - 2 x_{m,c} j_{n+1}(x_{m,c})}
{(n-1)j_n(x_{m,c}) - x_{m,c} j_{n+1}(x_{m,c})} 
- \frac{2n(n+1)[(1-n)j_n(x_{m,s})+x_{m,s} j_{n+1}(x_{m,s})]}
{(2n^2 - x_{m,s}^2 - 2)j_n(x_{m,s})+2x_{m,s} j_{n+1}(x_{m,s})} \right]^{-1},
\end{split}
\end{equation}
where $x_{m,c}=(c_0/c_c) x_m$ and $x_{m,s}=(c_0/c_s) x_m$.  

The solution for the rigid and movable sphere can be obtained by 
setting $c_c, c_s \rightarrow \infty$ in the previous discussion.
The solution for liquid spheres is achieved by letting $c_s \rightarrow 0$. 

\subsection{Second-order ultrasound scattering}
Before calculating the dynamic radiation force, we need to analyze the contribution
of the second-order velocity potential in Eq.~(\ref{f_force2}). 
The amplitude function $\hat{\phi}_{\Delta \omega}^{(2)}$ for an incident 
wave is calculated in the Appendix. 
In the pre-shock wave range, it is given by 
\begin{equation}
\label{phi_2} 
 \hat{\phi}^{(2)}_{\Delta \omega} = -\frac{\varepsilon v_0^2}{2 \Delta \omega} 
 je^{-j \Delta k z}.
\end{equation}
where  $\varepsilon = 1+B/(2A)$, $v_0$ is the peak amplitude of the velocity potential 
at the ultrasound source, and $\Delta k = \Delta \omega /c_0$. 
To simplify our analysis it was assumed that $\Delta k z \ll 1$.
If the difference frequency is about \unit[$10$]{kHz}, then the target should be around
\unit[$1$]{cm} of the ultrasound source.

Now, we need to solve the scattering problem for the second-order velocity potential.
This problem is similar to the linear scattering presented in Sec.~\ref{subsec:us_scat}. 
Hence, the total second-order velocity potential amplitude can be written 
in spherical polar coordinates as
\begin{equation}
\label{tot_phi_delta}
\hat{\phi}^{(2)}_{\Delta \omega} = -j\frac{\varepsilon v_0^2}{2 \Delta \omega} 
\sum_{n=0}^{\infty}(2n+1)(-j)^n [j_n(\Delta k r) + S_n h^{(2)}_n(\Delta k r)] 
P_n(\cos \theta), 
\end{equation}
where the scattering function $S_n$ is given through Eqs.~(\ref{S_mn}) and (\ref{D_mn}) by setting
$x_m = \Delta k r_0$.

According to Eq.~(\ref{f_force2}) the contribution of the second-order velocity potential is
given by integrating Eq.~(\ref{tot_phi_delta}) over the sphere surface. 
This contribution to the dynamic radiation force has only one component in the $z$-direction
given by
\begin{equation}
F_1 = -j 2\pi r_0^2 \rho_0 \Delta \omega \int_0^\pi \hat{\phi}^{(2)}_{\Delta \omega}(r_0) 
\sin \theta \cos \theta d\theta  =
- \pi r_0^2 E_0 \frac{4\varepsilon}{3} [j_1(\Delta k r_0) + S_1h_1^{(2)}(\Delta k r_0)],
\label{contrib_phi2}
\end{equation}
where $E_0 = (\rho_0 v_0^2)/2$ is ultrasound energy at the wave source.

\subsection{Dynamic radiation force function}
To calculate the dynamic radiation force we introduce the following 
variables $u_m=k_m r$ $(m=a,b)$ and $v=\cos \theta$. 
By symmetry considerations the dynamic radiation force on the sphere is only in the 
$z$-direction. 
By substituting Eq.~(\ref{tot_phi}) into Eq.~(\ref{f_force2}) leads to the amplitude of
the dynamic radiation force as  
\begin{equation}
\label{force_sphere}
\hat{f}_{\Delta \omega} =  F_0 + F_1 + F_2 + F_3 + F_4 + F_5,  
\end{equation}
where the amplitude functions are
\begin{eqnarray}
{F}_2 &=&  - \pi r_0^2 \rho_0 k_a k_b \label{a1}
\int_{-1}^{1}{\hat{\phi}_a(x_a,\nu) \hat{\phi}_b^{*}(x_b,\nu)v dv}, \\
{F}_3 &=& - \pi r_0^2 \rho_0 k_a k_b
\int_{-1}^{1}{\frac{\partial\hat{\phi}_a}{\partial u_a}\biggr|_{u_a=x_a} 
\frac{\partial \hat{\phi}_b^{*}}{\partial u_b}\biggr|_{u_b=x_b} v dv},\\
{F}_4 &=&  \pi \rho_0 
\int_{-1}^{1}{\frac{\partial\hat{\phi}_a}{\partial v}\biggr|_{u_a=x_a}  
\frac{\partial \hat{\phi}_b^{*}}{\partial v}\biggr|_{u_b=x_b} v(1-v^2) dv},\\
{F}_5 &=& - \pi r_0 \rho_0 
\int_{-1}^{1}{ \left( k_a \frac{\partial \hat{\phi}_b^{*}}{\partial v}\biggr|_{u_b=x_b}  
\frac{\partial\hat{\phi}_a}{\partial u_a}\biggr|_{u_a=x_a} 
+ k_b\frac{\partial\hat{\phi}_a}{\partial v} \biggr|_{u_a=x_a} 
\frac{\partial \hat{\phi}_b^{*}}{\partial u_b}\biggr|_{u_b=x_b} 
\right ) (1-v^2) dv}, \label{a4}
\end{eqnarray}
and $F_0$ and $F_1$ are given by Eqs.~(\ref{F_phi_1}) and (\ref{contrib_phi2}), respectively.

To obtain the amplitude functions of the dynamic radiation force in Eq.~(\ref{force_sphere}), 
we substitute Eq.~(\ref{tot_phi}) into Eqs.~(\ref{a1})-(\ref{a4}), which leads to
\begin{align}
\label{f2}
{F}_2 =& - \frac{2\pi r_0^2 E_{\Delta \omega}}{x_a x_b}  
\sum_{n=0}^\infty {(n+1)x_a x_b(R_{a,n} R_{b,n+1}^{*}+  R_{a,n+1} R_{b,n}^{*})}, \\
{F}_3 =& - \frac{2\pi r_0^2 E_{\Delta \omega}}{x_a x_b} 
\sum_{n=0}^\infty {(n+1) x_a x_b(R_{a,n}' {R'}_{b,n+1}^{*} + R_{a,n+1}' {R'}_{b,n}^{*})}, \\
{F}_4 =&  \frac{2\pi r_0^2 E_{\Delta \omega}}{x_a x_b}\sum_{n=0}^\infty { n(n+1)(n+2) 
(R_{a,n}R_{b,n+1}^* + R_{a,n+1}R_{b,n}^*) }, \\
\begin{split}
{F}_5 =& \frac{2\pi r_0^2 E_{\Delta \omega}}{x_a x_b} 
\sum_{n=0}^\infty  [ n(n+1) ( x_b R_{a,n} {R'}_{b,n+1}^{*} + x_a R_{a,n+1}' R_{b,n}^{*} ) \\
 &- (n+1)(n+2) ( x_a R_{a,n}' R_{b,n+1}^{*} + x_b R_{a,n+1} {R'}_{b,n}^{*} )], 
\label{amp4}
\end{split}
\end{align}
where $R_{m,n} = (-j)^n [j_n(x_m) + S_{m,n} h^{(2)}_n(x_m)]$ and 
$E_{\Delta \omega}= \rho_0 k_a k_b A_0^2$
is the difference frequency component of the ultrasound energy density. 

Let us focus on the contribution of $F_0$ to the dynamic radition force on the sphere.
The magnitude of the velocity particle of the incident plane wave is $p_0/(\rho_0 c_0)$,
where $p_0$ is the magnitude of the incident pressure.
The magnitude of the velocity particle inside the object volume variation 
$\delta V$ is smaller than its counter-part in the fluid.
Hence, from Eq.~(\ref{F_phi_1}) we have 
$|F_0| < \Delta \omega \delta M_{\text{max}} p_0/(\rho_0 c_0) $, 
where $\delta M_{\text{max}}$ is the maximum amount of fluid mass dislocated by the sphere 
vibration.
Measurements of the amplitude dislocation induced by the dynamic radiation force 
($\Delta f < \unit[1]{kHz}$, $f_0$ around \unit[$1$]{MHz}, and $p_0 < \unit[60]{kPa}$) on a stainless steel sphere of \unit[$1$]{mm} diameter in water, yielded results less than \unit[$1$]{$\mu$m}~\cite{chen04c}. 
Now, we compare the magnitude of $F_0$ and $F_2$. 
From Eq.~(\ref{f2}) we have 
\begin{equation*}
\frac{|F_0|}{|F_2|} < \frac{\rho_0 f_0^2 \Delta f r_0^2 \delta r}{c_0 p_0},
\end{equation*}
where $\delta r$ is the sphere radius variation. For the given condition of the 
experiment on measuring the dislocation amplitude of the sphere caused by the dynamic 
radiation force, the ratio $|F_0|/|F_1| \sim 10^{-2}$.
In this case or whenever $\delta r$ is negligible, only the components $F_1$ to $F_5$ are relevant to the radiation force 
formula~(\ref{force_sphere}).
 
One can show from Eq.~(\ref{rad_force_time_domain}) that the dynamic radiation force on the sphere in time-domain is
\begin{equation}
\mathbf{f}_{\Delta \omega}(t) 
= \pi r_0^2 E_{\Delta \omega} 
\Re \left\{ \hat{Y}_{\Delta \omega} e^{j \Delta \omega t} \right\} \mathbf{e}_z,
\end{equation}
where $\hat{Y}_{\Delta \omega}$ is the dynamic radiation force function given by
\begin{align}
\nonumber
\hat{Y}_{\Delta \omega} &= -\frac{2}{x_a x_b}
\sum_{n=0}^\infty (n+1)[ (x_a x_b-n(n+2))(R_{a,n} R_{b,n+1}^{*} + R_{a,n+1} R_{b,n}^{*})\\
& \quad + n (x_b R_{a,n} {R'}_{b,n+1}^{*} + x_a R_{a,n+1}' R_{b,n}^{*}) 
\nonumber
- (n+2)(x_a R_{a,n}' R_{b,n+1}^{*} + x_b R_{a,n+1} {R'}_{b,n}^{*})\\ 
& \quad + x_a x_b (R_{a,n}' {R'}_{b,n+1}^{*} + R_{a,n+1}' {R'}_{b,n}^{*})] - R_0,
\label{Yd}
\end{align}
where $R_0=\frac{4\varepsilon E_0}{3E_{\Delta \omega}} [j_1(\Delta k r_0) + S_1h_1^{(2)}(\Delta k r_0)]$.

The sphere is also subjected to a static radiation force, 
which is the sum of the force due to each ultrasound wave in the incident beam.
The static radiation force on a spherical target has been calculated by
Hasegawa~\cite{hasegawa69a}. 
The result reads
\begin{equation*}
\mathbf{f}_s  = \pi r_0^2 (E_a Y_a + E_b Y_b) \mathbf{e}_z, 
\end{equation*}
where $E_m = \rho_0 (k_mA_0)^2/2$ $(m=a,b)$. 
The quantity $Y_m$ is the radiation pressure function given by
\begin{equation}
\label{Yp}
Y_m =-\frac{4}{x_m^2} \sum_{n=0}^\infty (n+1) (\alpha_{m,n} + \alpha_{m,n+1} 
+ 2 \alpha_{m,n} \alpha_{m,n+1} + 2 \beta_{m,n} \beta_{m,n+1}),\quad m=a,b,
\end{equation}
where $\alpha_{m,n}$ and $\beta_{m,n}$ are the real and the imaginary parts 
of $S_{m,n}$, respectively. 
Moreover, when $\Delta \omega = 0$ then $\hat{Y}_{\Delta \omega} = Y_m$.

Finally, the total radiation force exerted on the sphere by the dual-frequency plane
wave is given by
\begin{equation}
\mathbf{f}_{\text{rad}}(t) = \pi r_0^2 \left( E_aY_a + E_b Y_b +
E_{\Delta \omega} 
\Re \left\{ \hat{Y}_{\Delta \omega} e^{j\Delta \omega t} \right\} \right) \mathbf{e}_z.
\end{equation}

\subsection{Numerical evaluation of the dynamic radiation force}
\label{subsec:numerical}
The dynamic radiation force function $\hat{Y}_{\Delta \omega}$ was evaluated
numerically in \textsc{Matlab} 6.5 (MathWorks, Inc.).
Two different materials were chosen in this evaluation: brass and stainless steel. 
The physical constants of the analyzed spheres are given in Table~\ref{parameters}. 
The surrounding fluid of the sphere was assumed to be water with density  
\unit[$\rho_0 = 1000$]{Kg/m$^3$}, compressional velocity \unit[$c_0 = 1500$]{m/s}.
The radius of the sphere is $r_0=\unit[0.5]{mm}$.

We are interested here in analyzing how the dynamic radiation force changes with
the center frequency $f_0$.
We evaluate the function $\hat{Y}_{\Delta \omega}$ as a function of $k_0r_0$ 
varying from $0.1$ to $10$. 
The difference frequency $\Delta f$ was fixed to $0$, $50$, \unit[100]{kHz}.
To assure that the center frequency $f_0$ is always positive, we assume that $f_0$ varies 
from \unit[$50$]{kHz} to \unit[$4.77$]{MHz}. 

\begin{table}[t]
\caption{Physical constants used in the calculation of the radiation force functions.}
\label{parameters}
\begin{center}
\begin{tabular}{lccc}
\hline
\hline
&&\multicolumn{2}{c}{Speed of sound}\\
\cline{3-4}
Material 
& Density
& Compressional
& Shear\\
&
Kg/m$^3$&
m/s&
m/s\\
\hline
Brass & 8100 & 3830 & 2050 \\
Stainless steel & 7670 & 5240 & 2978 \\
\hline
\hline
\end{tabular}
\end{center}
\end{table}

Before presenting the numerical evaluation results, let us take
a closer look to the contribution of the second-order velocity potential to the 
dynamic radiation force given in Eq.~(\ref{contrib_phi2}). 
For the frequency-range considered here, the energies $E_0$ and $E_{\Delta \omega}$
are about the same order of magnitude.
In this case, the numerical evaluation of $R_0$, see Eq.~(\ref{Yd}), 
for the chosen frequency-range has shown that the dimensionless amplitude of this quantity 
($10^{-3}$) is much smaller than the unit. 
In fact, for the frequency range used in the simulations we have 
$\Delta k r_0\ll 1$.
Therefore, we may neglect the contribution of $R_0$ to the dynamic radiation force
function given by Eq.~(\ref{Yd}) hereafter.

\begin{figure}[t]
  \centering
  \includegraphics[width=5.5in]{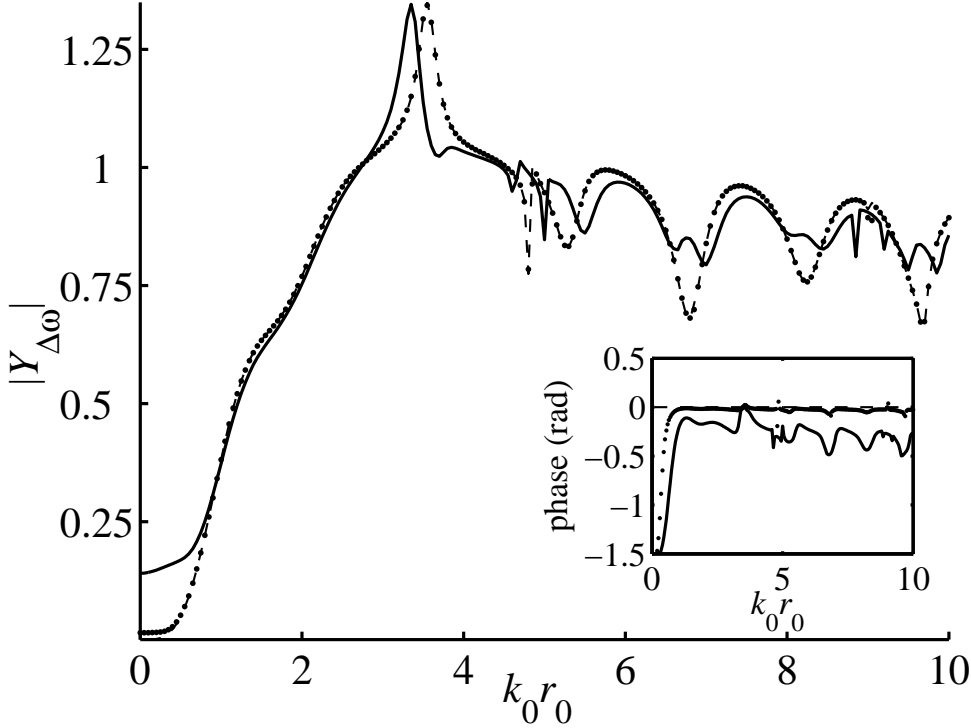}
  \caption{Dynamic radiation force function of the brass sphere.
  The inset plot the phase of  $\hat{Y}_{\Delta \omega}$. 
  Legend: dashed line (\unit[$0$]{kHz}), dotted line (\unit[$50$]{kHz}), 
  and solid line (\unit[$100$]{kHz}).}
  \label{fig:Yd_brass}
\end{figure}

\begin{figure}[t]
\centering
  \includegraphics[width=5.5in]{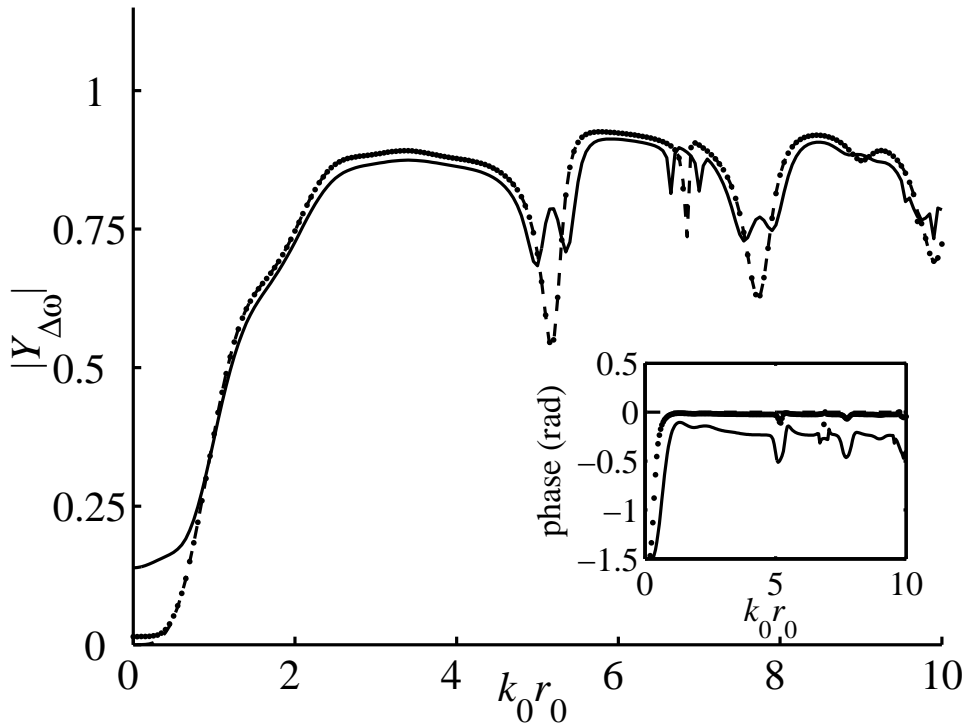}
  \caption{Dynamic radiation force function of the stainless steel sphere.
  The inset plot the phase of  $\hat{Y}_{\Delta \omega}$. 
  Legend: dashed line (\unit[$0$]{kHz}), dotted line (\unit[$50$]{kHz}), 
  and solid line (\unit[$100$]{kHz}).}
  \label{fig:Yd_steel}
\end{figure}

In Figs.~\ref{fig:Yd_brass} and~\ref{fig:Yd_steel}, we see the magnitude of the 
dynamic radiation force function $|\hat{Y}_{\Delta \omega}|$.
The inset of the figures shows the phase of $\hat{Y}_{\Delta \omega}$.
We can see that when $\Delta f = 0$ the function $|\hat{Y}_{\Delta \omega}|$ of both materials 
is equal to the radiation pressure function $Y_m$, as expected.
The dynamic radiation force function of both material exhibits a fluctuation pattern 
(dips and peaks) due to resonances of the ultrasound wave inside the sphere.
The fluctuations depend on resonances of the scattering function $S_{m,n}$,
which is related to the material parameters (density, compressional and shear speed of 
the wave). 
No significant changes in $\hat{Y}_{\Delta \omega}$ (magnitude and phase) is observed as the 
difference frequency varies from $0$  to \unit[$50$]{kHz}. 
Further, the phase remains practically constant with zero value. 
This occurs because at $\unit[$50$]{kHz}$ we have $\Delta k r_0 = 0.02$, which implies
$k_a r_0 \simeq k_b r_0$. Thus $\hat{Y}_{\Delta \omega}$ approaches to $Y_m$.

For the brass sphere, a prominent peak occurs in $|\hat{Y}_{\Delta \omega}|$ 
with $\Delta f = 0$ at $k_0r_0 = 3.55$, as seen in Fig.~\ref{fig:Yd_brass}. 
When the difference frequency is \unit[$100$]{kHz}, the peak change its position
to $3.27$ and the whole fluctuation pattern changes. 
However, the fluctuation form follows that one of $\Delta f = 0$ with smaller amplitudes. 
The phase also presents fluctuation whose amplitude is approximately
\unit[$\pi/6$]{rad}.

The function $|\hat{Y}_{\Delta \omega}|$ for the stainless steel sphere with $\Delta f= 0$ 
presents dips, as shown in Fig.~\ref{fig:Yd_steel}.  
The first dips occurs at $k_0r_0=5.17$. At \unit[$\Delta f = 100$]{kHz}, the fluctuations
in $\hat{Y}_{\Delta \omega}$ have a different pattern with smaller amplitude compared to the 
case of $\Delta f = 0$. 
The phase of $\hat{Y}_{\Delta \omega}$ is practically constant when
\unit[$\Delta f = 50$]{kHz}, except when $k_0r_0<1$. 
For \unit[$\Delta f=100$]{kHz}, the phase exhibits fluctuations with amplitude 
of about \unit[$\pi/6$]{rad}.
The phase fluctuations follows the pattern exhibited in the magnitude of 
$\hat{Y}_{\Delta \omega}$. 

\begin{figure}[t]
  \centering
  \includegraphics[width=5.5in]{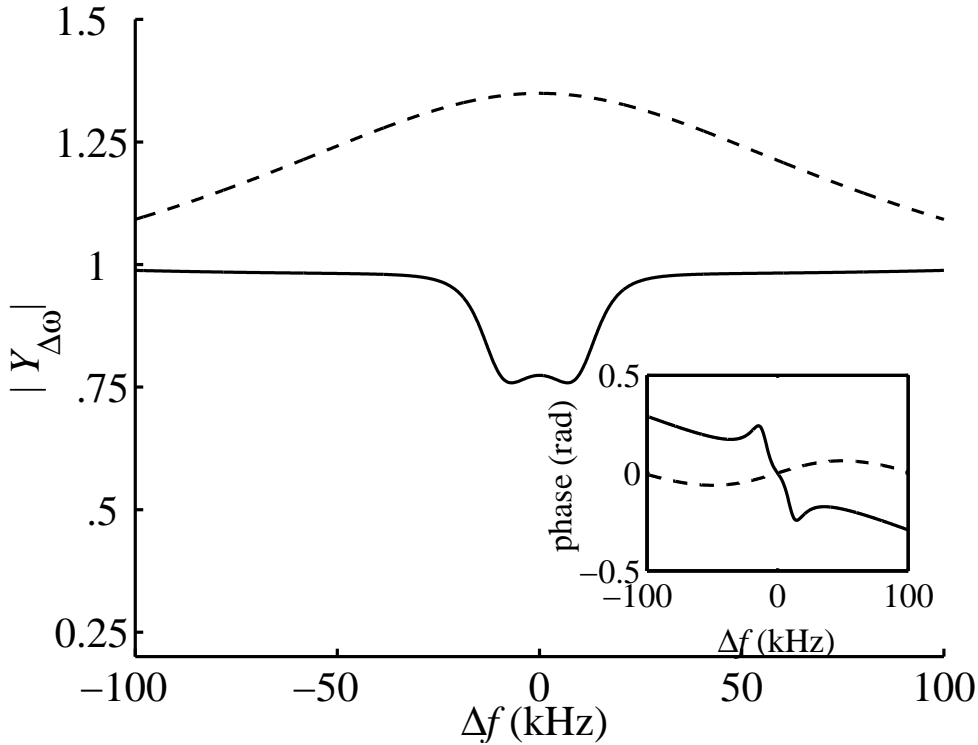}
  \caption{
  Dynamic radiation force function as $\Delta f$ varies for the brass sphere. 
  Legend: dashed line $(k_0r_0=3.55)$ and solid line $(k_0r_0=4.82)$.}
  \label{fig:Yd_df_brass}
\end{figure}

Now we analyze how the dynamic radiation force changes as $\Delta f$ varies.
This is particularly useful to show possible resonances on the dynamic radiation force
function of the spheres for a given center frequency $f_0$.
In Fig.~\ref{fig:Yd_df_brass}, the function $|\hat{Y}_{\Delta \omega}|$ is plotted 
as $\Delta f$ varies in from \unit[$-100$]{kHz} to \unit[$100$]{kHz}. 
The phase of $\hat{Y}_{\Delta \omega}$ is shown in the inset of the figure.
The frequency-range of $\Delta f$ was chosen to reveal symmetry properties of
$\hat{Y}_{\Delta \omega}$.
We fixed $k_0r_0$ to $3.55$ and $4.82$ for the dashed and solid lines,
respectively. 
These values correspond to the first peak and dip, respectively, in Fig.~\ref{fig:Yd_brass}.
In both cases the $|\hat{Y}_{\Delta \omega}|$ is symmetric. 
The phase of $\hat{Y}_{\Delta \omega}$ is antisymmetric.
Both magnitude and phase of $\hat{Y}_{\Delta \omega}$ change shape considerably
as $k_0r_0$ changes.
The concavities of $|\hat{Y}_{\Delta \omega}|$ in Fig.~\ref{fig:Yd_df_brass}
follow those shown in Fig.~\ref{fig:Yd_brass} for the resonance points
$3.55$ (peak) and $4.82$ (dip). 

\begin{figure}[t]
  \centering
  \includegraphics[width=5.5in]{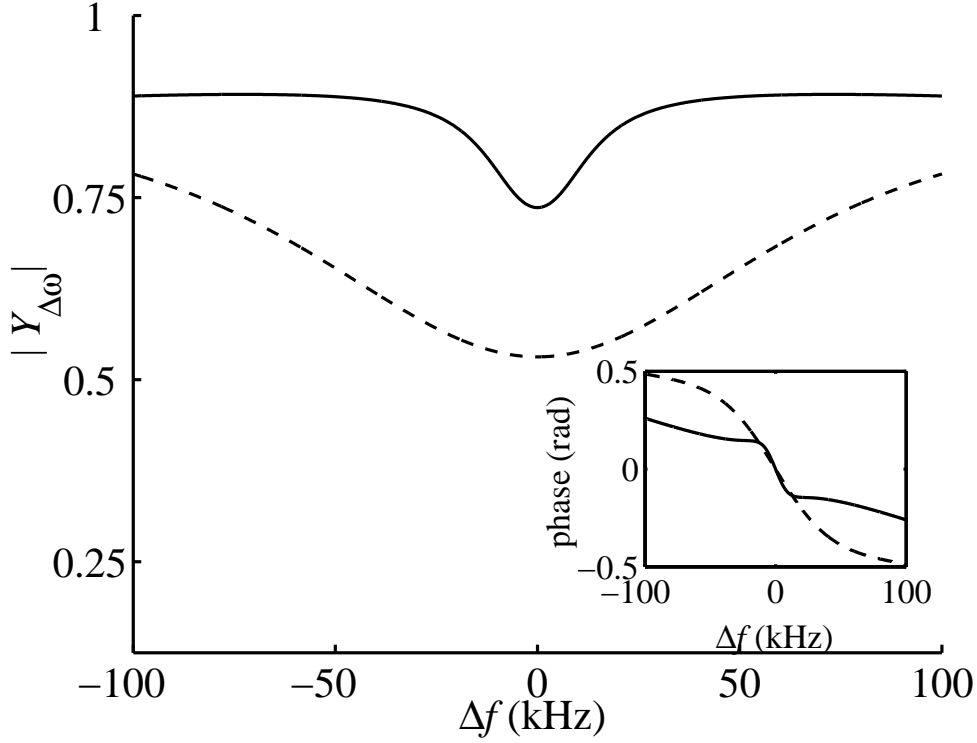}
  \caption{
  Dynamic radiation force function as $\Delta f$ varies for the stainless steel sphere. 
  Legend: dashed line $(k_0r_0=5.17)$ and solid line $(k_0r_0=6.85)$.}
  \label{fig:Yd_df_steel}
\end{figure}

The plot of the function $\hat{Y}_{\Delta \omega}$ of the stainless steel sphere
as $\Delta f$ varies from \unit[$-100$]{kHz} to \unit[$100$]{kHz} is shown
in Fig.~\ref{fig:Yd_df_steel}. 
The inset of the figure plots the phase of $|\hat{Y}_{\Delta \omega}|$.
The quantity $k_0r_0$ was fixed at the first and second dips, which corresponds
to $5.17$ and $6.85$ for the dashed and solid lines, respectively.
The function $|\hat{Y}_{\Delta \omega}|$ for both values of $k_0r_0$ exhibited 
a symmetric concave pattern. 
The phase showed an antisymmetric pattern in both cases.  
The concavities of $|\hat{Y}_{\Delta \omega}|$ seen in Fig.~\ref{fig:Yd_df_steel}
follow those presented in Fig.~\ref{fig:Yd_steel} for points
$5.17$ and $6.85$.

\begin{figure}[t]
  \centering
  \includegraphics[width=5.5in]{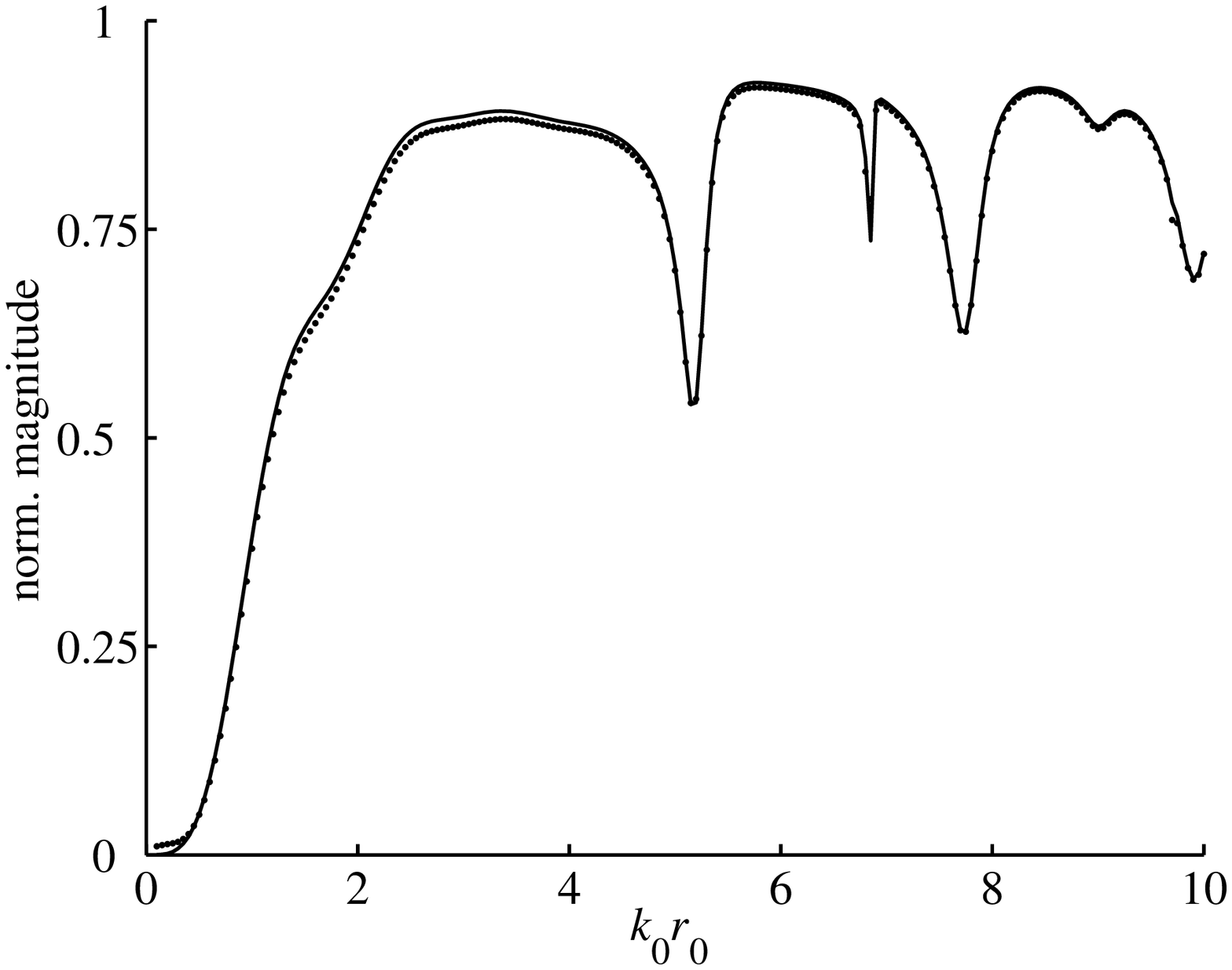}
  \caption{Comparison on the magnitude of the static and the dynamic ultrasound radiation
  force on the stainless steel sphere with $\Delta f = \unit[10]{kHz}$. Legend: dotted line 
  represents 
$(k_b/k_a)|\hat{Y}_{\Delta \omega}|$ and solid line stands for $Y_a + (k_b/k_a)^2 Y_b$. 
}
  \label{fig:comp_mag}
\end{figure}

The comparison of the normalized amplitude of the static and the dynamic radiation 
forces on the stainless sphere as the center frequency varies can be seen in 
Fig.~\ref{fig:comp_mag}. 
The radiation force functions $Y_a$ and $Y_b$ are given by Eq.~(\ref{Yp}).
The difference frequency was fixed to \unit[$10$]{kHz}.
The amplitudes are normalized by the highest ultrasound energy density $E_a$ times
the cross-section area of the sphere. 
In this figure, the solid line stands for the amplitude of the static radiation force,
$Y_a + (k_b/k_a)^2 Y_b$. 
The dotted line corresponds to $(k_b/k_a)|\hat{Y}_{\Delta \omega}|$.
The magnitude of the both radiation forces are about the same. 
Though we changed the difference frequency in a range up to \unit[$100$]{kHz}, the
magnitude of the static and the dynamic radiation force remained approximately the same.

\section{Conclusions}
\label{sec:discussion}
We have presented a theory to calculate the dynamic ultrasound radiation force
exerted on an object by a dual-frequency CW ultrasound beam in lossless fluids.
The theory is valid for beams with any spatial distribution.
The amplitude of the induced vibration by the dynamic radiation 
force on the object was assumed to be much smaller than its characteristic
dimensions.
No assumptions have been made on geometric shape of the object.
A formula for the dynamic radiation force was obtained -- see Eq.~(\ref{f_force2})-- in terms
of the first- and second-order velocity potentials.
The dependence of the dynamic radiation force with the nonlinear parameter 
$B/A$ of the medium was analyzed.

We have calculated the dynamic radiation force exerted on a solid 
sphere by a dual-frequency CW plane wave in water. 
The dynamic radiation force is proportional to the cross-section area of the sphere, 
the dynamic ultrasound energy, and the dynamic radiation force function. 
The contribution of the first-order velocity potential to the radiation force,
accounted by Eq.~(\ref{force_sphere}), was neglected because we considered that 
the dislocation of the sphere is very small.
The contribution of the medium nonlinearity to the dynamic radiation force is negligible 
if $\Delta k r_0 \ll 1$ in a weak nonlinear medium.
However, it may become more significant in strongly nonlinear mediums or when
$\Delta k r_0 \sim 1$.
Numerical evaluation of the dynamic radiation force function revealed a fluctuation pattern 
as the center frequency varies.
The fluctuations are similar to those present in the static radiation force function.
Analysis of the static and the dynamic radiation force on the sphere has shown that they 
have approximately the same magnitude.
Experimental measurements of the magnitude of the static and the dynamic radiation force on a 
stainless steel sphere have confirmed this prediction~\cite{chen04c}.
%

In conclusion, the presented dynamic radiation force formula~(\ref{f_force2}) generalizes 
Yosioka's formula~\cite{yosioka55a}, which stands only for the static radiation force.
The dynamic radiation force formula can be extended for a multi-frequency
CW ultrasound beam. 
This is particularly useful to study pulsed radiation force in which the incident 
ultrasound beam can be expanded in time as a Fourier series. 
	
\section*{Acknowledgments}
This work was partially supported by grant DCR013.2003-FAPEAL/CNPq (Brazil).

\appendix
\section{Second-order velocity potential}
To calculate the contribution of the second-order velocity potential to 
the dynamic radiation force on the sphere, we consider the lossless Burger's 
equation 
\begin{equation*}
\frac{\partial v}{\partial z} - \frac{\varepsilon}{c_0^2}v
\frac{\partial v}{\partial \tau} = 0,
\end{equation*}
where $v$ is the velocity particle in the $z$-direction, $\varepsilon = 1+B/(2A)$,  
and $\tau = t-z/c_0$ is the retarded time.
The source wave form is given by
\begin{equation*}
v(0,\tau) = v_0 [\sin (\omega_a\tau)  +  \sin (\omega_b \tau)], 
\end{equation*}
where $v_0$ is the peak amplitude of the velocity particle at the wave source $(z=0)$.
Hence, the second-order velocity particle at the difference frequency 
in the pre-shock wave range is given by~\cite{fenlon72a}
\begin{equation*}
v^{(2)}_{\Delta \omega} \simeq
- \frac{\varepsilon v_0^2}{2 c_0}
\Delta k z \sin(\Delta \omega t - \Delta kz), 
\end{equation*}
This approximated solution is valid for $\epsilon v_0 \Delta k z / c_0 \ll 1$.
We obtain the second-order velocity potential at the difference frequency
by integrating this equation over $z$.
Thus, in complex notation, the amplitude of the velocity potential at $\Delta \omega$
is
\begin{equation*}
\hat{\phi}^{(2)}_{\Delta \omega} = \frac{\varepsilon v_0^2}{2 \Delta \omega}
(\Delta k z - j) e^{-j \Delta k z}.
\end{equation*}
Notice that the time-dependent integration constant was dropped because
it evaluates zero in the closed surface integral of Eq.~(\ref{f_force2}).

\newpage
\bibliographystyle{apsrev}


\end{document}